\documentclass[acmsmall]{acmart}
\AtBeginDocument{%
  }

\setcopyright{acmlicensed}
\copyrightyear{2018}
\acmYear{2018}
\acmDOI{XXXXXXX.XXXXXXX}

\acmJournal{JACM}
\acmVolume{37}
\acmNumber{4}
\acmArticle{111}
\acmMonth{8}

\usepackage{enumitem}

\begin{document}

\definecolor{exercisebgblue}{rgb}{0,  .69,  .941}
\definecolor{pastelviolet}{rgb}{.81,  .82,  .97}

\newcommand{\abhik}[1]{\textcolor{magenta}{\textbf{Abhik:} #1}}
\newcommand{\rangeet}[1]{\textcolor{blue}{\textbf{Rangeet:} #1}}
\newcommand{\yintong}[1]{\textcolor{red}{\textbf{Yintong:} #1}}
\title{Towards Risk-free AI Agent Deployment}


\author{Yintong Huo}
\affiliation{%
  \institution{Singapore Management University}
  \city{Singapore}
  \country{Singapore}}
\email{ythuo@smu.edu.sg}

\author{Rangeet Pan}
\affiliation{%
  \institution{IBM T.J. Watson Research Center}
  \city{Yorktown Heights, NY}
  \country{USA}}
\email{rangeet.pan@ibm.com}

\author{Abhik Roychoudhury}
\affiliation{%
  \institution{National University of Singapore}
  \city{Singapore}
  \country{Singapore}}
\email{abhik@nus.edu.sg}






\renewcommand{\shortauthors}{Trovato et al.}

\begin{abstract}
LLM-based agents are rapidly moving from research prototypes into the core business processes of organizations, but these agents pose deployment risks to security, compliance, and functionality. In this article, we argue that risk-free deployment must be grounded in the agent's trajectory: the recorded sequence of reasoning steps, tool invocations, and environmental observations. Trajectories are available for any agent, and many failures are visible only in the trajectory. To make agents deployable and sustainable, we advocate agent testing and debugging as a systematic research direction for detecting and mitigating these risks. This article begins with the challenges of testing agents, including the oracle problem, non-determinism, trajectory validation, and the absence of adequacy metrics. We then turn to debugging agents, from automated failure attribution to repair and self-evolution. We distill these directions into a practical deployment-readiness checklist covering the full deployment lifecycle. Finally, we identify open problems, i.e., formal adequacy metrics, root-cause attribution over long-horizon trajectories, and the reliability of self-evolving agents, that the community must address to enable trustworthy agent deployment.

\end{abstract}





\received{20 February 2007}
\received[revised]{12 March 2009}
\received[accepted]{5 June 2009}

\maketitle

\section{Industrial Transformation}

Agents have had a significant impact not only on computing, but also on society at large. Agents provide autonomous decision-making via the use of tools, memory, and planning. It is thus quite significant that many manual or human-guided activities in computing, such as coding, code review, and system validation, are being taken over by agents.

In software engineering per se, interest in agents is a few years old at this point. The main interest started with the design of coding agents, as it was found that Large Language Models (LLMs) are capable of vibe-coding and generating code for natural language requirements. Soon afterwards, there were suggestions for invoking tools such as command-line utilities (bash tools) and program analysis tools to support code generation, program repair, and feature-addition tasks, i.e., generative tasks. A commentary on coding agents and the role of trust in using coding agents appears in a recent Communications of the ACM article \cite{cacm26}. 

While coding agents from software engineering have evoked significant interest in the broader generative AI and agentic AI communities, they are {\em only} one class of agents. The broader automation possibilities from agentic AI are not restricted to coding agents. Any organization with a significant business process or workflow can obtain potential productivity gains by automating parts of that workflow. As a simple example, a bank can automate part of its operational workflow by automating the loan approval task via a curated AI agent. However, such agentification of organizational workflows is currently ad hoc. If loan approval is done by an agent, it can introduce security, compliance, or functionality-breaking risks into the rest of the bank's processes. Currently, such risks of agentification are not being actively studied or understood; instead, specific components of an organization's workflow are being agentified in isolation.

\begin{figure}[tbp]
    \centering
    \includegraphics[width=\textwidth]{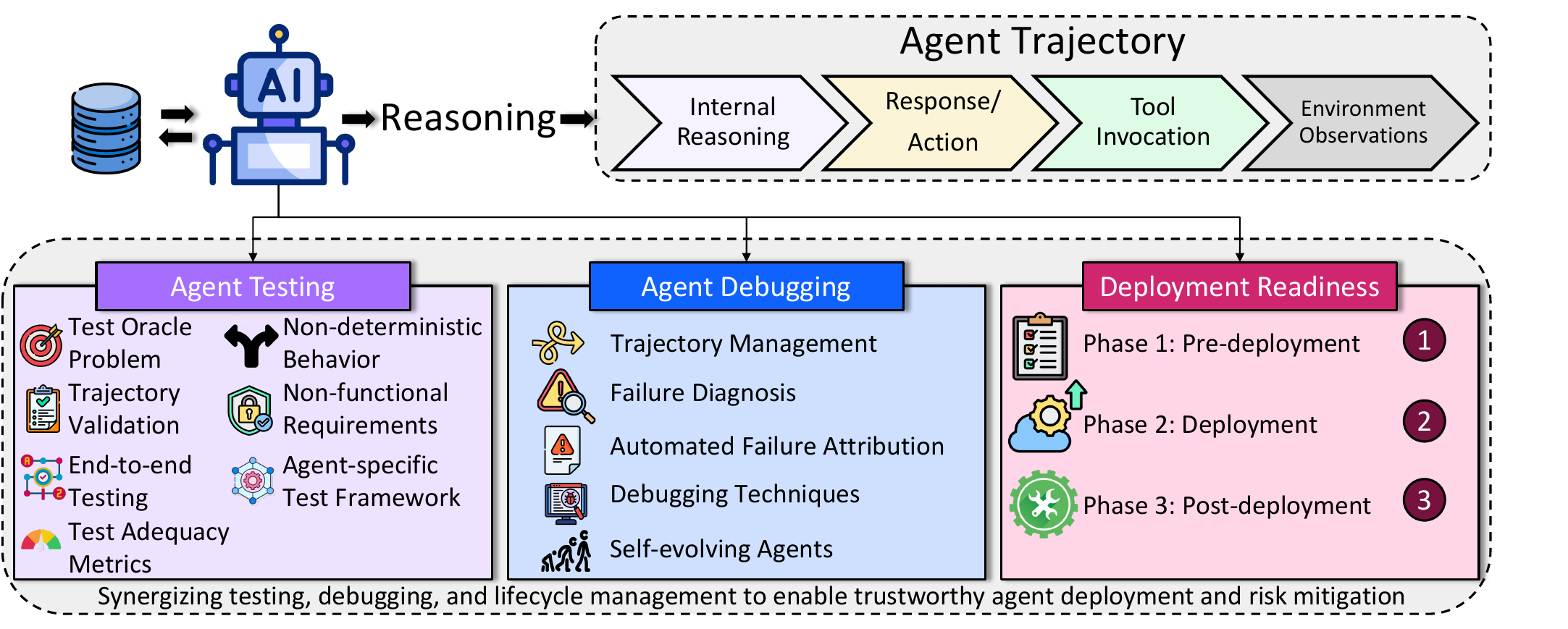}
    \caption{The overview of this article. The roadmap to risk-free deployment includes agent testing, agent debugging, and deployment readiness checklists.}
    \label{fig:overview}
\end{figure}

The aim of this article is to present a suite of ideas, techniques, and technologies that prompt us to consider the risk-free deployment of agents. To achieve such risk-free deployment, one of the first steps is to systematically test and debug AI agents. Such testing and debugging methodologies should be applicable to any agent, not only coding agents. Furthermore, ideally, the testing and debugging methodologies should work with agents using both open-weight and closed models. This is possible if the methodologies are based on agent trajectories. As shown in Fig.~\ref{fig:overview}, in this paper, we thus present a suite of agent testing and debugging methodologies to enable the risk-free deployment of agents in organizations. We also discuss the impact of agentification on the rest of the organizational workflow. The techniques and ideas presented can provide an evaluation layer that supports the risk-free agentification of an organizational workflow.

\section{Agents as a Reactive System}
Traditional software is deterministic: given the same input, the software produces the same output. Agents, by contrast, are fundamentally reactive and non-deterministic systems. They perceive an environment, update their internal state based on observations and reasoning, and generate actions that change the environment, creating a feedback loop that can persist for dozens or hundreds of steps. Many agents realize this loop through a ReAct (Reason-and-Act) architecture, which interleaves reasoning with tool-based actions. The ReAct architecture introduces profound testing and debugging challenges.

An agentic system's execution can be represented as a \textit{trajectory}, i.e., a sequential record of the agent's thinking process, tool invocations, environmental observations, and responses. Each step in a trajectory contains four critical types of information: the agent's internal reasoning, its response or proposed action, the concrete actions taken (such as code modifications or tool use), and the resulting observations from the environment. This trajectory captures the agent's decision-making process and provides insight into its behavior. Many agent failures are visible only in the trajectory, making the trajectory itself a primary resource for testing and debugging~\cite{wang2026agent}.

The reactive nature of agents makes their failure modes different from traditional software. A single flawed reasoning step early in an agent's execution can cascade through subsequent steps, leading to compounded errors that manifest only at the trajectory's end. For example, an agent might misunderstand a user's instruction in its first step, adopting an incorrect implementation that goes undetected until it falls into an infinite loop. This implicit relationship between cause and effect makes agents harder to debug than classical systems, and it obscures which step to hold responsible for an observed failure.

Furthermore, agents operate under continuous uncertainty about their environment. Tool outputs contain noise, incomplete information, and unexpected results, and agent actions may produce side effects on external services and the state. The sheer length of agentic trajectories (for example, often exceeding 40 steps in project issue resolution) introduces additional strain on the agent's reasoning capability. Research shows that even state-of-the-art LLMs struggle to maintain reasoning quality when processing extended contexts, compounding the challenges of both testing and debugging.

\section{Challenges in Testing Agents}
\label{sec:challenges}

Testing LLM-based agent applications introduces challenges that go beyond those encountered in conventional software testing. These challenges stem from the various properties of agentic systems: non-deterministic behavior, multi-step reasoning, tool use, interaction with external environments, etc. Together, these properties complicate how developers define correctness, observe failures, measure coverage, and automate regression testing. In this section, we summarize the key challenges that arise when testing agentic systems.

\subsection{Test Oracle Problem}
\label{subsec:oracle}

The test oracle problem---determining whether an observed behavior is correct---is especially acute in LLM-based agent applications. In conventional software, testers can often specify an expected output for a fixed input and compare the observed result against it. In agentic systems, however, correctness is often \emph{distributional rather than point-valued}. The same query may produce different, yet equally acceptable, responses across executions. Acceptability may also depend on context, phrasing, intermediate reasoning paths, and the persona or role of the end user, such as a product manager versus a technical lead. As a result, assertions must be flexible enough to admit legitimate variation while still being precise enough to detect real failures.

\subsection{Testing Non-deterministic Behavior}
\label{subsec:nondeterminism}

Non-determinism is one of the most fundamental obstacles to reliable agent testing. Because LLM inference is stochastic, repeated executions of the same test may produce different final outputs, intermediate tool calls, or reasoning paths. This undermines traditional testing assumptions, where a fixed input is expected to produce a fixed and reproducible output. A test may pass in one run and fail in another without any change to the system under test, producing behavior analogous to flaky tests in conventional software~\cite{dutta2020detecting}.

Non-determinism also complicates debugging and test isolation. When a failure occurs, developers may be unable to determine whether it reflects a genuine defect, a sampling artifact, an environmental condition, or an interaction among these factors. Reproducing failures requires controlling, replaying, or at least logging the stochastic elements of execution, but current testing frameworks provide limited support for doing so.

Several approaches from software testing offer useful foundations. Property-based testing~\cite{hughes2007quickcheck} and metamorphic testing~\cite{Chen1998MetamorphicTesting} shift the focus from exact output equivalence to relational properties that should hold across executions. Techniques from probabilistic program testing~\cite{ Borges2015DistributionAwareSampling, Luckow2014ExactApproximatePSE} and model checking~\cite{Mardziel2013ProbabilisticAbstractInterpretation} provide formal foundations for reasoning about stochastic systems. Recent work has begun adapting these ideas to agents: ToolFuzz~\cite{milev2025toolfuzz} fuzzes tool runtimes, ChainFuzzer~\cite{wu2026chainfuzzer} targets multi-tool workflows, and agentic property-based testing~\cite{maaz2025agentic} explores bug finding across LLM-powered components. However, a comprehensive framework for managing non-determinism across prompts, models, tools, memory, and external environments remains absent.

\subsection{Testing Trajectories}
\label{subsec:trajectories}

Agent behavior is not fully captured by final outputs. Many failures arise from the \emph{trajectory} of execution: the sequence of reasoning steps, tool selections, delegation decisions, intermediate states, and environment transitions that lead to a final result. Testing these trajectories is therefore one of the most distinctive challenges in agent testing.

Observability platforms such as LangSmith, LangFuse, Arize AI, and Maxim AI expose execution traces, intermediate decisions, tool calls, and environment transitions. However, these artifacts are typically external to the main test suite and are not well integrated with standard testing frameworks such as JUnit, PyTest, Unittest, etc. Consequently, even when developers inspect trajectories during debugging, those checks are rarely encoded as automated, reusable assertions. This creates a gap between what can be observed and what is systematically validated during routine development.

Bridging this gap requires testing infrastructure that supports first-class assertions over trajectories, including tool-selection paths, delegation patterns, reasoning traces, and multi-step task execution~\cite{kim2025beyond}. Such assertions may need to operate at different levels of granularity: individual tool calls, sub-task plans, full execution traces, or resulting environment states. Formalizing what constitutes an acceptable trajectory, and defining adequacy criteria based on trajectory coverage rather than only code coverage, remain important open problems.

\subsection{Testing Non-functional Requirements}
\label{subsec:nfr}

Non-functional requirements (NFRs)---including security, performance, resilience, privacy, and policy compliance---are critical for agentic systems because agents increasingly interact with external tools, sensitive data, and real-world environments. Yet NFR testing remains underrepresented in practice. Empirical evidence from open-source agent projects~\cite{pan2026tangent} shows that only a small fraction of tests, roughly 7--8\%, target NFRs at all. Among those tests, most focus on security scenarios, while performance, memory, and resilience receive substantially less attention.

This scarcity reflects several compounding difficulties. Performance and latency are difficult to evaluate without realistic workloads and deployed environments, which are often unavailable during unit testing. Security properties, such as prompt-injection resistance, PII anonymization, and privilege-escalation prevention, require adversarial inputs that are difficult to construct systematically. Resilience testing, which evaluates behavior under partial failures such as tool timeouts, schema drift, or unreachable services, requires fault-injection capabilities that current agent frameworks rarely expose.

Established software engineering techniques provide a useful starting point. Fault injection~\cite{arlat1990fault}, chaos engineering~\cite{basiri2016chaos}, and metamorphic testing~\cite{chen2018metamorphic} are well-established approaches for evaluating behavior under failures and adversarial conditions. Recent work has begun extending these approaches to agents: MAST~\cite{cemri2025multi} catalogs behavioral failure modes in multi-agent execution traces, ReliabilityBench~\cite{gupta2026reliabilitybench} applies chaos engineering to single-agent systems using infrastructure-level faults, and MAS-FIRE~\cite{jia2026mas} defines a fault taxonomy for multi-agent systems. However, prior work has not yet provided a unified fault taxonomy or injection mechanism that spans the full agentic stack, including tool integration, memory, inter-agent communication, security, and fault propagation across architectural layers.

\subsection{End-to-end Testing}
\label{subsec:e2e}

End-to-end (E2E) testing exercises a complete flow from user input through all system layers to the final outcome. For agentic systems, E2E testing is especially important because correctness often depends on the composition of agents, tools, memory, external services, and environment side effects across a full task. At the same time, E2E testing is difficult to perform in practice. It typically requires a deployed or near-deployed environment, realistic user inputs, and mechanisms for observing and asserting on outcomes that may span multiple turns, tool invocations, and external state changes.

A central difficulty is distinguishing integration testing from true E2E testing when the application is not fully deployed. E2E tests also face the full force of the oracle problem (Section~\ref{subsec:oracle}) and non-determinism (Section~\ref{subsec:nondeterminism}), because both are amplified when the entire system is exercised together. Validating environment side effects, such as files written, database entries modified, or external services called, adds further complexity.

Recent benchmarks such as $\tau$-bench~\cite{yao2024tau} and OSWorld~\cite{xie2024osworld} move toward E2E evaluation by assessing resulting system state rather than only final outputs. However, systematic support for E2E testing of agent workflows in development settings remains limited. Developers still lack practical tools for constructing realistic scenarios, controlling external dependencies, validating side effects, and replaying failures with sufficient determinism for regression testing.

\subsection{Test Adequacy Metrics}
\label{subsec:adequacy}

Measuring the adequacy of a test suite---how much of the system has been exercised and how well---is a long-standing concern in software testing. Conventional criteria such as statement coverage, branch coverage, and mutation score provide well-defined measures for traditional software. These criteria, however, do not capture the higher-level abstractions that characterize agentic systems, including tool-selection paths, reasoning traces, delegation patterns, multi-step task execution, memory interactions, and environment state transitions.

In practice, no widely accepted adequacy standard exists for agent testing. Practitioners instead rely on ad hoc approaches such as edge-case coverage, domain-driven test design, trajectory sampling, query clustering, and acceptable failure thresholds---all of which are project-specific judgments rather than formal criteria~\cite{pan2026tangent}. This creates uncertainty about when a test suite is sufficiently comprehensive and makes it difficult to compare testing practices across projects.

Developing adequacy metrics for agentic systems likely requires moving beyond code coverage toward behavioral and interaction coverage. Examples include coverage over tool-use patterns, prompt and input distributions, execution trajectories, memory states, delegation structures, and environment-side effects. These metrics must also account for stochastic behavior, since a single execution may not adequately characterize an agent's behavior under a given scenario.

\subsection{Agent-specific Test Frameworks}
\label{subsec:framework}

Despite the rapid growth of LLM-based agent applications, the tooling and infrastructure for testing them remain immature. General-purpose frameworks such as PyTest and Unittest were designed for conventional software and do not natively support assertions over execution trajectories, delegation behavior, reasoning steps, non-deterministic outputs, or environment side effects. Although observability platforms expose many of these artifacts, they are often separate from the test suite and are not designed for automated, regression-style validation.

Agent-specific test infrastructure is therefore needed to treat trajectories, tool calls, intermediate reasoning states, memory interactions, and environment transitions as first-class test targets. Such infrastructure should provide: (1) reusable test doubles that simulate not only tool execution behavior but also the effects of tool descriptions on agent reasoning~\cite{hasan2026model}; (2) distributional assertions for validating stochastic behavior across repeated executions; (3) mechanisms for replaying execution traces with sufficient determinism for regression testing; and (4) integration with observability data so that trajectory-level assertions can be embedded within standard testing workflows. Developing this infrastructure, grounded in a rigorous theory of agent testability that extends classical notions of controllability and observability~\cite{freedman1991testability, voas2002software} to the agentic setting, is a key prerequisite for systematic and rigorous testing of LLM-based agent applications.

\section{Challenges in Debugging Agents}
Another central requirement for risk-free agent deployment is the ability to debug agents. In classical software engineering, debugging proceeds by inspecting the execution state of a program using execution traces. 
The analogous artifact for an LLM-based agent is its trajectory: the recorded sequence of observations, reasoning steps, and tool calls. Because trajectories can be produced by any agent, building the debugging layer on top of them yields methodologies that generalize across coding agents, enterprise workflow agents, and any other deployment class an organization might consider.

We organize the body of work on trajectory-based agent debugging into three areas: (1) trajectory management, where trajectories must be captured, structured, and stored so that they can be inspected and reasoned over, (2) failure analysis that detects a failure and attributes it to a specific component or step, and (3) the diagnosis have to be turned into an actual fix, either by correcting the ongoing run or by ensuring the agent does not repeat the mistake in the future. We discuss each area in depth below.

\subsection{Capturing Agent Trajectories}
Capturing agent trajectories is the foundational step for any post-hoc analysis, failure diagnosis, or debugging effort. However, trajectory capture is not simply logging; it requires a systematic approach to recording, standardizing, and managing execution data. The need for robust trajectory management becomes significant with the scale of modern agentic runs. A single SWE-bench coding task may involve dozens of file reads, shell commands, and LLM reasoning steps. An enterprise workflow agent processing a multi-document loan application might execute hundreds of API calls across several sub-agents. Storing, indexing, and retrieving this data with enough fidelity to support debugging is a research problem.

\textbf{Standardized trajectory format.} 
Raw agent logs are unstructured flat text and lack standards, whereas failure analysis tools have to operate on a standard representation. On the infrastructure side, the practical implementation layer is served by open-source tools such as Langfuse, to capture LLM calls, tool invocations, and latency as structured, queryable traces. At the standardization level, OpenTelemetry defines cross-framework semantic conventions for LLM and agent tracing, with adoptions already appearing in the industry. Graphectory~\cite{liu2026process} is a directed graph constructed automatically from raw trajectories in which nodes represent agent actions and edges encode both temporal sequencing and structural navigation.

\textbf{Noise and context management.} 
Raw trajectories are noisy: a long-horizon task can exceed one hundred thousand tokens, most irrelevant to any given failure. The community has converged on memory distillation as the primary solution. Agent Workflow Memory~\cite{wang2025agent} induces compact, reusable workflow subroutines from prior trajectories, improving task accuracy through smarter trace utilization. ReasoningBank~\cite{ouyang2025reasoningbank} distills generalizable reasoning strategies via a self-judging mechanism, establishing experience scaling as a new performance dimension.

\subsection{Failure diagnosis}
Failure analysis asks: given a trajectory, did the agent fail, why did it fail, and at what point did the failure become inevitable? This area encompasses failure characterization and attribution.  The difficulty of this problem should not be underestimated. Unlike a traditional program that throws an exception at the point of failure, an LLM-based agent can fail silently and finally lead to wrong or unsafe output. The failure may be owed to an early bad decision whose consequences only become visible many steps later. 

\textbf{Characterizing Agent Failures.} Before failures can be detected or attributed, they must be understood empirically. Failure characterization studies help practitioners identify which instrumentation to deploy, which guardrails to implement, and where human oversight is most needed. The most systematic single-agent effort is Bouzenia et al.~\cite{bouzenia2025understanding}, identifies recurring behavioral anti-patterns across 120 trajectories, including incoherent reasoning chains, failure to integrate tool feedback, and action-repetition loops. Liu et al.~\cite{liu2026evaluating} isolate plan deviation as a distinct failure class across 10K trajectories, finding that a poorly specified plan inflicts more damage than no plan at all. For multi-agent systems, MAST~\cite{cemri2026multi} summarizes 14 failure modes across 1,600+ traces, highlighting that coordination breakdowns and system design deficiencies are as harmful as the individual model error.

\textbf{Automated Failure Attribution.} Even when a failure is detected, attributing it to the right step in a long trajectory is a challenging research problem because of its long and noisy causal chains. A wrong decision at step 7 may not surface until later steps fail with test suites. Naive approaches that inspect only the final few steps, or treat all steps as equally likely candidates, fall short in uncovering the failure root causes.
The field has developed with early benchmarks and several strategies. 
Who\&When~\cite{zhang2025agent} benchmarks automated attribution directly, introducing three baseline strategies—all-at-once, step-by-step, and binary-search attribution—and finding that even frontier models achieve only a miserable step-level localization accuracy.
AgentRx~\cite{barke2026agentrx} advances step localization through constraint synthesis: it prompts an LLM to generate behavioral constraints a correct execution should satisfy, evaluates each constraint against each trajectory step, and passes the resulting step-indexed validation log to a judge—substantially reducing the reasoning burden over raw traces. 
RootSE~\cite{wang2026trajaudit} is the first failure diagnosis benchmark on agentic coding trajectories, characterized by noise from redundant program structure and verbose code. It follows up a diagnostic solution where the failure-irrelevant content is pre-folded via pattern matching, and an investigator iteratively retrieves necessary execution information on demand.

\subsection{Debugging: End-to-end validation}
The focus closes the loop from diagnosis to improvement. Existing methods can be categorized into two directions: Intra-task repair uses the failure diagnosis to correct the agent by re-executing. Inter-task self-evolution persists the lessons learned across tasks, building up a growing base of experience to improve itself over time.  For organizational deployment, end-to-end debugging is where the practical value of the entire pipeline materializes. An agent that can detect its own mistakes and recover from them is safer to deploy in a consequential workflow. An agent that learns from its failures over time progressively reduces its error rate, which is the necessary condition for sustainable automation.

\textbf{Intra-task debugging.} 
The immediate challenge for intra-task debugging is translating a failure diagnosis into an effective correction. If the diagnosis is too vague, the agent has insufficient information to do better on a retry; if the feedback is too specific, it may not generalize to related failures. The quality of the corrective signal is the primary determinant of whether repair succeeds.
Reflexion~\cite{shinn2024reflexion} established the foundational mechanism - an agent writes a verbal self-reflection after each failed attempt and stores it in a memory buffer - reaching an advanced performance without any model weight updates. AgentDebug~\cite{zhu2025llm} integrates the step-level attribution with targeted feedback injection, re-executing the agent from the identified fault point with error-grounded context. Its recovery gains demonstrate that attribution precision is a direct predictor of repair quality. Wink~\cite{nanda2026wink} classifies live trajectories' misbehaviors across pre-defined failure categories, and intervenes in agent runtime behavior with course-correction guidance.

\textbf{Self-evolving through trajectory learning.}
Self-evolving agents go beyond one-shot repair; they extract reusable knowledge from past trajectories and accumulate it in persistent memory or skill libraries that shape future behavior. The key insight is that trajectories are not merely execution logs but a form of implicit knowledge: by extracting and reusing what has worked and what has not, an agent can improve across tasks over time without retraining. An early survey of this line of work argues that self-evolution through trajectory learning is the defining long-term challenge for autonomous agent systems~\cite{gao2025survey}. Specifically, SE-Agent~\cite{guo2026se} realizes this by revising, recombining, and refining prior reasoning traces, yielding consistent improvements on multi-step coding tasks. SkillRL~\cite{xia2026skillrl} formalizes the distillation step: a teacher model converts raw trajectories into reusable skills, enabling the agent to generalize learned behaviors to new tasks rather than re-solving from scratch. Trace2Skill~\cite{ni2026trace2skill} provides direct trajectory-to-skill distillation: a set of sub-agents extracts trajectory-level lessons through inductive reasoning and consolidates them into a unified, reusable skill directory for future use. 
Ensuring the quality of an evolving skill library remains the central open challenge: incompetent skills introduce noise and contradiction into agent execution, ultimately risking performance degradation.

\section{Deployment Readiness: A Checklist for Organizations}

The testing and debugging methodologies discussed above establish the research foundation for safer agent deployment.
This section provides a checklist to help organizations assess their readiness to deploy an agent into core business processes.
The checklist is organized around three deployment phases: pre-deployment, deployment, and post-deployment.

\subsection{Pre-Deployment Readiness}
\begin{itemize}[leftmargin=*]
    \item \textbf{Have you established a trajectory capture system?}
    Without systematic trajectory logging, it is impossible to debug failures after deployment and to learn from agent behavior at scale. 

    \textit{How:} Use infrastructure like LangFuse, Langsmith, or OpenTelemetry to record all agent reasoning steps, tool calls, and environment observations in a queryable format. Trajectory data should be retained for at least the duration of your rollout plan, and indexed by task, date, and outcome.
    \item \textbf{Have you characterized failure modes specific to your domain?} The failure patterns discovered in one domain (e.g., improper reasoning chains in coding) may not be the primary failure modes in yours (e.g., policy compliance checking). Your testing strategy should target your most likely failure modes first.

    \textit{How:} Conduct a sampled trajectory review with domain experts and the agent team to identify recurring behavioral anti-patterns. Document at least critical failure scenarios as in ~\cite{bouzenia2025understanding} and ~\cite{liu2026evaluating}.
    
    \item \textbf{Have you defined acceptance criteria that your agent must meet?} Acceptance criteria translate your business requirements into measurable assertions. This provides an objective basis for determining whether the agent is ready for deployment.

    \textit{How:} Define criteria across multiple dimensions: functional correctness, safety, and compliance rate. Frame criteria in terms of trajectory properties or distributional outcomes rather than single-run outputs.
    
    \item \textbf{Have you executed an end-to-end test in a staging environment that mirrors production?} Unit and integration tests exercise components in isolation. End-to-end tests run the full workflow, including all tools, external services, and data flows. Failures discovered in staging are orders of magnitude cheaper to fix than failures discovered in production.

    \textit{How:} Run your agent through realistic, complete workflows in an environment as close to production as possible. Validate both the final output and procedural actions (files written, database records modified, external services called). Document latency, error rates, and recovery behavior. 
    
    \item \textbf{Have you established human oversight and rollback procedures?} Even well-tested agents fail in production under conditions not anticipated during development. Human oversight ensures that high-stakes decisions are reviewed, and rollback procedures ensure that failures do not cascade.

    \textit{How:} Define which agent decisions require human review before being executed, which require review after the fact, and which are fully autonomous. 
\end{itemize}

\subsection{Early Deployment and Monitoring}

\begin{itemize}[leftmargin=*]
    \item \textbf{Are you monitoring the execution process in trajectory?} Many agent failures are implicit in the final output but are visible in the trajectory (incorrect reasoning chains, failure to use available tools, or infinite loops). Trajectory monitoring catches failures before they cause business damage.

    \textit{How:} Set up automated alerts on trajectory properties: unusual tool-call sequences, repeated failed steps, or reasoning traces that deviate from expected patterns. Sample trajectories regularly for manual inspection by domain experts.

    \item \textbf{Have you established a feedback loop for capturing user-reported failures?} Users will encounter edge cases and contexts that your test suite did not anticipate. A structured process for capturing and diagnosing these failures ensures that real-world experience feeds back into improvements.

    \textit{How:} Create a reporting mechanism for users to flag unexpected behavior. Correlate each report with the corresponding trajectories and task environments. Prioritize reports by frequency and business impact, and route high-priority failures to your debugging team with full trajectory context attached.
    
    \item \textbf{Are you executing intervention on failed trajectories?} Many can be fixed by correcting the agent mid-execution or by providing targeted feedback that helps the agent recover. Intra-task debugging is faster and less risky than redeployment.

    \textit{How:} When a task failure is detected, before rolling back or escalating, attempt one or more recovery strategies: ask the agent to reflect on its error and retry~\cite{shinn2024reflexion}, re-execute from the fault point with corrected context~\cite{wang2026trajaudit}, or inject course-correction guidance~\cite{nanda2026wink}. Log the outcome and the triggering error for later analysis.

\end{itemize}

\subsection{Sustained Deployment and Improvement}

\begin{itemize}[leftmargin=*]
    \item \textbf{Are you storing lessons learned from trajectories into reusable knowledge?} Over time, agents should become more reliable through learned experience. Self-evolution through trajectory learning allows agents to internalize patterns and avoid repeating mistakes.

    \textit{How:} Periodically review successful and failed trajectories to extract patterns. Consolidate these into reusable skills or decision heuristics that the agent can reference in future runs. Update the agent's system prompt, memory, or tool set to incorporate these lessons. Validate that the agent improves on similar tasks in subsequent runs.
    
    \item \textbf{Have you measured and reported your adequacy of testing?} Stakeholders need to understand beyond test acceptance metrics, for example, how comprehensively we have exercised the agent and its failure modes?

    \textit{How:} Report metrics beyond pass/fail rate, such as trajectory diversity, failure-mode coverage, consistencies, and distributional coverage across different scenarios or prompts.

\end{itemize}

\section{Open Research Questions for the Community}
While trajectory-based testing and debugging provide a systematic foundation for agent deployment, three fundamental research challenges remain open:

\textbf{Formal Test Adequacy Metrics for Agentic Systems.}
Software testers have long relied on metrics like statement coverage and branch coverage. But these do not work for agents. What matters for agents is different: tool selection diversity, reasoning quality, trajectory variation, and memory interactions.
Today, practitioners rely on intuition to choose edge cases based on domain knowledge. Therefore, there exists a gap in understanding when testing is enough. A rigorous, formal adequacy metric for agents is needed. Such a metric must account for non-determinism and multi-step behavior. It would let practitioners answer the fundamental question: is my test suite comprehensive enough? It would also let researchers compare testing methodologies fairly.

\textbf{Root-cause attribution in Long-horizon Trajectories.}
When a traditional program fails, we see an exception and investigate it. When an agent fails, the diagnosis is much harder. In a trajectory spanning 50 to 200 steps, a wrong decision at step 7 may only become visible at step 50. By then, many other decisions have cascaded onto it.
Current approaches treat attribution as a search problem. Systems like AgentRx~\cite{barke2026agentrx} and RootSE~\cite{wang2026trajaudit} show promise. Yet even the best models achieve only modest accuracy in pinpointing the actual failure point. The field needs a capable framework for causal reasoning over noisy, stochastic trajectories. With better attribution, we can build debugging tools that practitioners can trust.

\textbf{Safety and Trustworthiness of Agents.}
Agents that learn from their own trajectories show real promise. They accumulate skills over time and improve themselves without retraining. But it also introduces new risks. Learned skills can be wrong. They can contradict each other. Agent behavior can drift away from what the organization intended. When an agent changes its own behavior, who audits that change? In this direction, the fundamental questions are still open. How do we verify that a skill library maintains safety invariants? How do we detect when learned skills degrade before they break production? When should humans intervene in an autonomous learning loop?

\section{Conclusion}
The transition from human-guided workflows to agentic automation is underway across industries, but deploying an agent into a consequential business process without the means to audit it is a critical risk. This article has argued that the path to risk-free deployment runs through the observable agent trajectory: a model-agnostic artifact for building systematic testing methodologies and a debugging pipeline for fixing issues.
The deployment-readiness checklist we presented translates these research advances into concrete organizational practice across the deployment lifecycle. We also identify several open research questions for the community, including formal adequacy criteria for stochastic multi-step systems, root-cause attribution in long-horizon trajectories, and mechanisms to keep self-evolving agents aligned with organizational intent. Addressing these challenges will require sustained collaboration among the software engineering and AI communities. 

\bibliographystyle{ACM-Reference-Format}
\bibliography{software}

\end{document}